\documentclass[usenatbib]{mn2e}
\usepackage{graphicx}

\title[Effect of molecular contamination in Orion A]{The JCMT Gould Belt Survey: the effect of molecular contamination in SCUBA-2 observations of Orion A}
\author[S. Coud\'{e} et al.]{S. Coud\'{e}$^{1}$\thanks{E-mail:
coude@astro.umontreal.ca}, P. Bastien$^{1}$, H. Kirk$^{2}$, D. Johnstone$^{2,3,4}$, E. Drabek-Maunder$^{5}$,\newauthor 
S. Graves$^{6,7,8}$, J. Hatchell$^{9}$, E. L. Chapin$^{4}$, A. G. Gibb$^{10}$, B. Matthews$^{2,4}$ and \newauthor the JCMT Gould Belt Survey Team\thanks{See JCMT Gould Belt Survey membership list in Appendix A}\\
$^{1}$Universit\'e de Montr\'eal, Centre de Recherche en Astrophysique du Qu\'ebec et d\'epartement de physique,\\ C.P. 6128, succ. centre-ville, Montr\'eal, QC, H3C 3J7, Canada\\
$^{2}$NRC Herzberg Astronomy and Astrophysics, 5071 West Saanich Rd, Victoria, BC, V9E 2E7, Canada\\
$^{3}$Joint Astronomy Centre, 660 N. A`oh\={o}k\={u} Place, University Park, Hilo, Hawaii 96720, USA\\
$^{4}$Department of Physics and Astronomy, University of Victoria, Victoria, BC, V8P 1A1, Canada\\
$^{5}$Imperial College London, Blackett Laboratory, Prince Consort Rd, London SW7 2BB, UK\\
$^{6}$Astrophysics Group, Cavendish Laboratory, J J Thomson Avenue, Cambridge, CB3 0HE, UK\\
$^{7}$Kavli Institute for Cosmology, Institute of Astronomy, University of Cambridge, Madingley Road, Cambridge, CB3 0HA, UK\\
$^{8}$East Asian Observatory, 660 N. A`oh\={}k\={u} Place, University Park, Hilo, Hawaii 96720, USA\\
$^{9}$Physics and Astronomy, University of Exeter, Stocker Road, Exeter EX4 4QL, UK\\
$^{10}$Department of Physics and Astronomy, University of British Columbia, 6224 Agricultural Road, Vancouver, BC, V6T 1Z1}
\begin{document}

\date{Accepted 2015 December 24. Received 2015 December 4; in original form 2015 August 21}

\pagerange{\pageref{firstpage}--\pageref{lastpage}} \pubyear{2015}

\maketitle

\label{firstpage}

\begin{abstract}
Thermal emission from cold dust grains in giant molecular clouds can be used to probe the physical properties, such as density, temperature and emissivity in star-forming regions. We present the SCUBA-2 shared-risk observations at 450 $\mu$m and 850 $\mu$m of the Orion A molecular cloud complex taken at the James Clerk Maxwell Telescope (JCMT). Previous studies showed that molecular emission lines can contribute significantly to the measured fluxes in those continuum bands. We use the HARP $^{12}$CO J=3-2 integrated intensity map for Orion A in order to evaluate the molecular line contamination and its effects on the SCUBA-2 maps. With the corrected fluxes, we have obtained a new spectral index $\alpha$ map for the thermal emission of dust in the well-known integral-shaped filament. Furthermore, we compare a sample of 33 sources, selected over the Orion A molecular cloud complex for their high $^{12}$CO J=3-2 line contamination, to 27 previously identified clumps in OMC-4. This allows us to quantify the effect of line contamination on the ratio of 850 $\mu$m to 450 $\mu$m flux densities and how it modifies the deduced spectral index of emissivity $\beta$ for the dust grains. We also show that at least one Spitzer-identified protostellar core in OMC-5 has a $^{12}$CO J=3-2 contamination level of $16 \%$. Furthermore, we find the strongest contamination level ($44 \%$) towards a young star with disk near OMC-2. This work is part of the JCMT Gould Belt Legacy Survey.
\end{abstract}


\section{Introduction}

Although interstellar dust grains represent only a small fraction ($\approx 1 \%$) of the total mass in molecular clouds, they are important sites for the production of molecular hydrogen (H$_{2}$) and other complex molecules. These molecules play a role in the formation and the chemistry of pre-stellar cores \citep{bib:30,bib:31}. Therefore, studying the physical properties of interstellar dust, such as temperature and emissivity, should inform us about the processes leading to the birth of stars and their stellar systems. Specifically, the emissivity represents the efficiency at which interstellar dust grains re-emit as thermal emission in the far infrared the starlight they absorbed. The emissivity spectral index $\beta$ is a parameter related to the composition and size distribution of the dust mixture, which can often be characterized by a power law \citep{bib:10}. With the current tools at our disposal to probe molecular clouds in the infrared, submillimetre and radio wavelengths, it is possible to achieve a more complete description of interstellar dust in regions of star formation.

The Gould Belt is a ring-like configuration of nearby O-type stars of $\approx 350$ pc diameter centered about 200 pc from the Sun \citep{bib:40}. It is associated with a number of active star formation regions such as the Orion Nebula and NGC 1333. The JCMT Gould Belt Survey \citep{bib:02} is an international effort to expand the study of cold interstellar matter in these regions with the help of SCUBA-2, a continuum instrument sensitive to thermal emission from dust grains \citep{bib:12}, and HARP, an heterodyne instrument useful for studying molecular line emission \citep{bib:13}.

The Orion A molecular cloud is a well known star formation region located at a distance between 400 and 500 pc from our solar system \citep{bib:03}. Along with Orion B, it is one of two giant molecular clouds in the Orion superbubble \citep{bib:01}. It contains the famous Orion Nebula (Messier 42) and a collection of smaller molecular complexes. Even though Orion objects have been observed extensively for decades, many questions remain unanswered. For example, it is still unclear which mechanism,turbulence or magnetism, is the most effective at regulating star formation in parts of the cloud \citep{bib:04}.

In this work, we present the SCUBA-2 shared risk observations of Orion A taken in February 2010 at the JCMT. The maps at 450 $\mu$m and 850 $\mu$m are used to evaluate the spectral index of emissivity in different regions of the Orion molecular cloud complex. We know from previous studies \citep{bib:05,bib:06,bib:28} that molecular contamination can be particularly important at these wavelengths and in this region. In order to evaluate its effects on the determination of the dust grain emissivity, we use HARP integrated intensity maps from \citet{bib:32} of the contaminating $^{12}$CO J=3-2 emission.

This paper is divided in four main sections. First, we describe a simple method to evaluate the spectral index of emissivity with SCUBA-2 observations. Then we present the observations, followed by the results. Finally, we discuss the importance of quantifying molecular contamination in the context of understanding star formation in Orion.

\section{Method}

As mentioned in the previous section, the emissivity $\kappa_\nu$ (cm$^{2}$ g$^{-1}$) at a given frequency $\nu$ (Hz) represents the efficiency at which dust grains re-emit as thermal emission the energy they absorbed at higher frequencies. This parameter is directly related to the size distribution and physical properties of dust mixtures. Since thermal emission from cold dust grains dominates the continuum in the submillimetre regime, the emissivity can be approximated from SCUBA-2 observations, assuming an optically thin medium, by using a simple power-law \citep{bib:07}. If $\kappa_o$ is the emissivity at a reference frequency $\nu_o$, the spectral index of emissivity $\beta$ is defined by: \begin{equation}
\label{eq:04}
\kappa_\nu = \kappa_o \left(\frac{\nu}{\nu_o}\right)^\beta.
\end{equation} A more detailed look at interstellar dust emissivity models is presented by \citet{bib:08}, but our simplified approach still provides a useful estimation of the size distribution and composition of the dust grains \citep{bib:09, bib:10}.

\begin{figure}
\includegraphics[width=84mm]{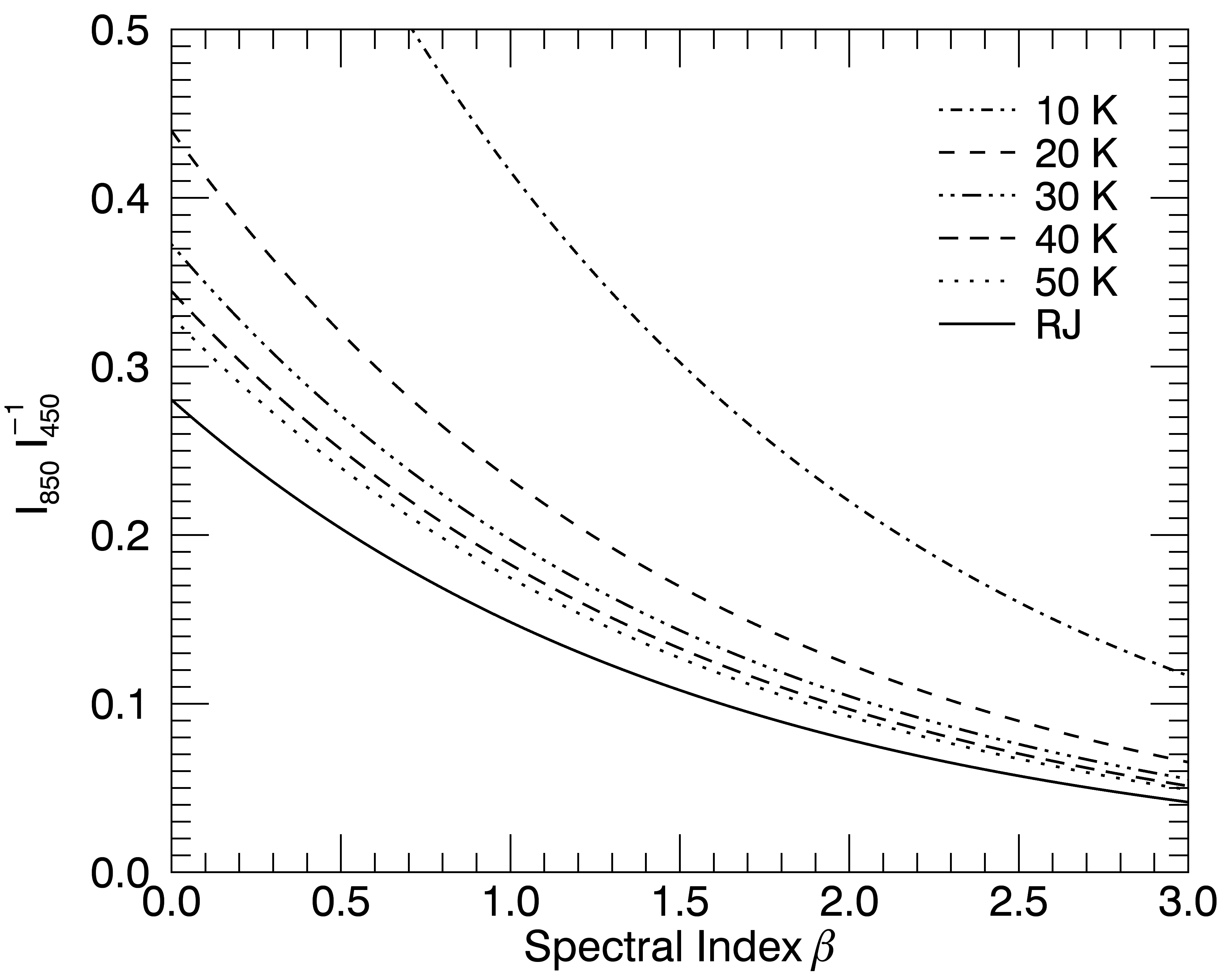}
\caption{Effect of temperature on the emissivity spectral index $\beta$. For a given temperature, the theoretical ratio of flux densities ($I_{850}$ $I_{450}^{-1}$) is plotted as a function of the spectral index, as defined from Equation \ref{eq:01}, for an ideal case where there is no molecular contamination ($C_{\lambda} =0$ $\forall\;\lambda$). The solid line represents the temperature-independent case, obtained with the Rayleigh-Jeans approximation for dust thermal emission.}
\label{fig:01}
\end{figure}

If dust grains are in thermal equilibrium with the radiation field, which is a reasonable assumption for large enough grains in molecular clouds \citep{bib:33,bib:30}, we can suppose that their emission follows a modified form of Planck's law $B_\nu$ (erg s$^{-1}$ cm$^{-2}$ Hz$^{-1}$ sr$^{-1}$) for a blackbody at a temperature $T_d$ (K). Modeling the 3D dust distribution in star-forming regions from only a handful of observations is, however, far from trivial \citep{bib:11}. In order to simplify the analysis included in this paper, we assume that the properties of dust grains are identical along any given line of sight. This permits the use of the dust column density $N_d$ (g cm$^{-2}$) to represent the distribution in the plane of the sky. We also introduce the fraction $C_\nu$ of contamination at a given frequency. This umbrella factor represents the total contribution to the measured intensity $I_\nu$ from sources other than the dust's thermal emission. From these assumptions, we can write the relationship between the total measured intensity $I_\nu$ and the dust thermal emission at a given frequency as: \begin{equation}
\label{eq:01}
I_\nu (1-C_{\nu}) = N_d \kappa_\nu(\beta) B_\nu(T_d).
\end{equation} Although this equation is defined as a flux density in units of frequency ($10^{23}$ Jansky sr$^{-1}$), the notation used from this point on will use wavelengths $\lambda$ ($\mu$m) instead to relate with the language commonly used for far-infrared astronomy. This will simplify the discussion without modifying the units used to quantify SCUBA-2 observations. We can now define the spectral index $\alpha$ as the slope of continuum thermal emission between two measured flux densities: 

\begin{equation}
\label{eq:02}
\frac{I_{\lambda_1} (1-C_{\lambda_1})}{I_{\lambda_2}(1-C_{\lambda_2})} = \left(\frac{\lambda_2}{\lambda_1}\right)^{\alpha}.
\end{equation}

If we define $\Delta\alpha$ as the difference between spectral indices $\alpha$ after and before correction for molecular contamination, we obtain this simple relation from Equation \ref{eq:02}:

\begin{equation}
\label{eq:05}
\Delta\alpha = \ln{\left(\frac{1 - C_{\lambda_1}}{1 - C_{\lambda_2}}\right)} \left[ \ln{\left(\frac{\lambda_2}{\lambda_1}\right)} \right]^{-1}.
\end{equation} 

The spectral index $\alpha$ is the measurable quantity from which we evaluate the emissivity spectral index $\beta$. If we combine Equations \ref{eq:01} and \ref{eq:02} at any given temperature $T_d$ and assume the dust emissivity from Equation \ref{eq:04}, we find that there exists a direct relationship between these spectral indices:

\begin{equation}
\label{eq:03}
\beta = \alpha - \ln{\left(\frac{B_{\lambda_1}(T_d)}{B_{\lambda_2}(T_d)}\right)} \left[ \ln{\left(\frac{\lambda_2}{\lambda_1}\right)} \right]^{-1}.
\end{equation}

For this two-wavelength method, we know from Equation \ref{eq:03} that any effect contamination has on the measured spectral index $\alpha$ is going to propagate linearly to the deduced emissivity spectral index $\beta$. Furthermore, any difference calculated for one of the spectral indices is going to be identical for the other ($\Delta\alpha = \Delta\beta$) for a given dust temperature $T_d$. This means that it is possible to evaluate the effect of molecular contamination on the emissivity spectral index $\beta$ with no information about the column density $N_d$ or the average temperature $T_d$ if it is assumed fixed.

For this study, we use SCUBA-2 maps of Orion A at 850 $\mu$m and 450 $\mu$m. We define the measured flux densities per unit of telescope beam area as $S_{850}$ and $S_{450}$ (Jansky beam$^{-1}$) respectively. If those flux densities are properly calibrated, normalized and beam-matched (see Section 3), they can be directly substituted for $I_{\lambda_1}$ and $I_{\lambda_2}$ in Equation \ref{eq:02} without any loss of information.

In the Rayleigh-Jeans approximation ($h\nu \ll kT_d$), the flux density $B_\nu$ in Equations \ref{eq:01} and \ref{eq:03} can be written as $B_\nu(T_d) = 2 (\nu/c)^2 k T_d$, where $k$ is Boltzmann's constant (erg K$^{-1}$) and $c$ is the speed of light (cm s$^{-1}$). This allows for a temperature-independent determination of the spectral index $\beta$ through a simple ratio of flux densities. In this regime, Equation \ref{eq:03} is simplified as follows: $\beta = \alpha - 2$. It is, however, true only in regions with high dust temperatures ($T_d > 50$ K), which is not the case in most prestellar cores \citep[$T_d \approx$ 10 - 15 K, see][]{bib:09}. The influence of temperature on the emissivity spectral index $\beta$ is shown in Figure \ref{fig:01}. It includes the specific case of the Rayleigh-Jeans approximation, without including contamination, as well as the ones calculated from the more general Planck law for a given series of temperatures (10 K to 50 K). From Figure \ref{fig:01}, it is clear that the main purpose of the Rayleigh-Jeans approximation is to establish a lower limit on the spectral index $\beta$ by directly measuring the spectral index $\alpha$ introduced in Equation \ref{eq:02}, without any prior assumption about temperature or density.

In order to evaluate the effect of molecular contamination on the spectral index of emissivity, we postulate that the fraction $C_{850}$ from Equation \ref{eq:02} comes entirely from $^{12}$CO J=3-2 rotational line emission. We use a molecular emission map converted from HARP spectroscopic data using the technique presented by \citet{bib:06}. It is important to note that this conversion depends on the effective transmission of the telescope, and therefore varies with the atmospheric conditions during the observations. Although it is not expected to be the only contributing factor, this $^{12}$CO J=3-2 molecular line is the most likely culprit of molecular contamination at 850 $\mu$m as it is the brightest and most common emission line over the entire region except in OMC-1's bright core. Away from hot cores, the $^{12}$CO J=3-2 line can be 5 to 10 times brighter in the 850 $\mu$m band than all other molecular lines combined \citep[see][]{bib:05}.

Similarly, we would expect the main driver of molecular contamination at 450 $\mu$m to be the emission from the $^{12}$CO J=6-5 rotational line. \citet{bib:06} extrapolated from $^{12}$CO J=3-2 observations that the molecular contamination $C_{450}$ is, in most cases, unlikely to be significant when compared to the dust thermal emission at this wavelength. This can be verified in OMC-1 using the $^{12}$CO J=6-5 integrated main-beam brightness temperature map from \citet{bib:51}. With the conversions factors from \citet{bib:06}, it is possible to estimate the $^{12}$CO J=6-5 contamination toward some key regions in our 450 $\mu$m observations: Orion KL, Orion South and the Orion bar. We estimate the contamination to be $< 1\%$, $< 1\%$ and $< 5\%$ of the total 450 $\mu$m flux density for these regions respectively. These estimates are likely to be upper limits since they do not take into account instrumental differences, such as sensitivity to large-scale emission, between CHAMP$^+$ \citep{bib:51}, HARP and SCUBA-2. It is therefore reasonable, at least for OMC-1, to assume the $^{12}$CO J=6-5 line contamination to be negligible for SCUBA-2 observations at 450 $\mu$m. Since other molecular lines in that observation band are generally much weaker than the $^{12}$CO J=6-5 line, except for hot cores \citep{bib:51},we limit this study only to the effects of contamination in the 850 $\mu$m band. The coefficient $C_{850}$ can only decrease the 850 to 450 $\mu$m flux ratio, therefore the spectral index of emissivity will be \textit{underestimated} if contamination from molecular emission is not taken into account. This may be particularly important in regions of strong molecular emission relative to the dust continuum, such as a low dust column density region containing molecular outflows \citep{bib:41}.

\section{Observations}

\begin{figure*}
\begin{minipage}{175mm}
\includegraphics[width=175mm]{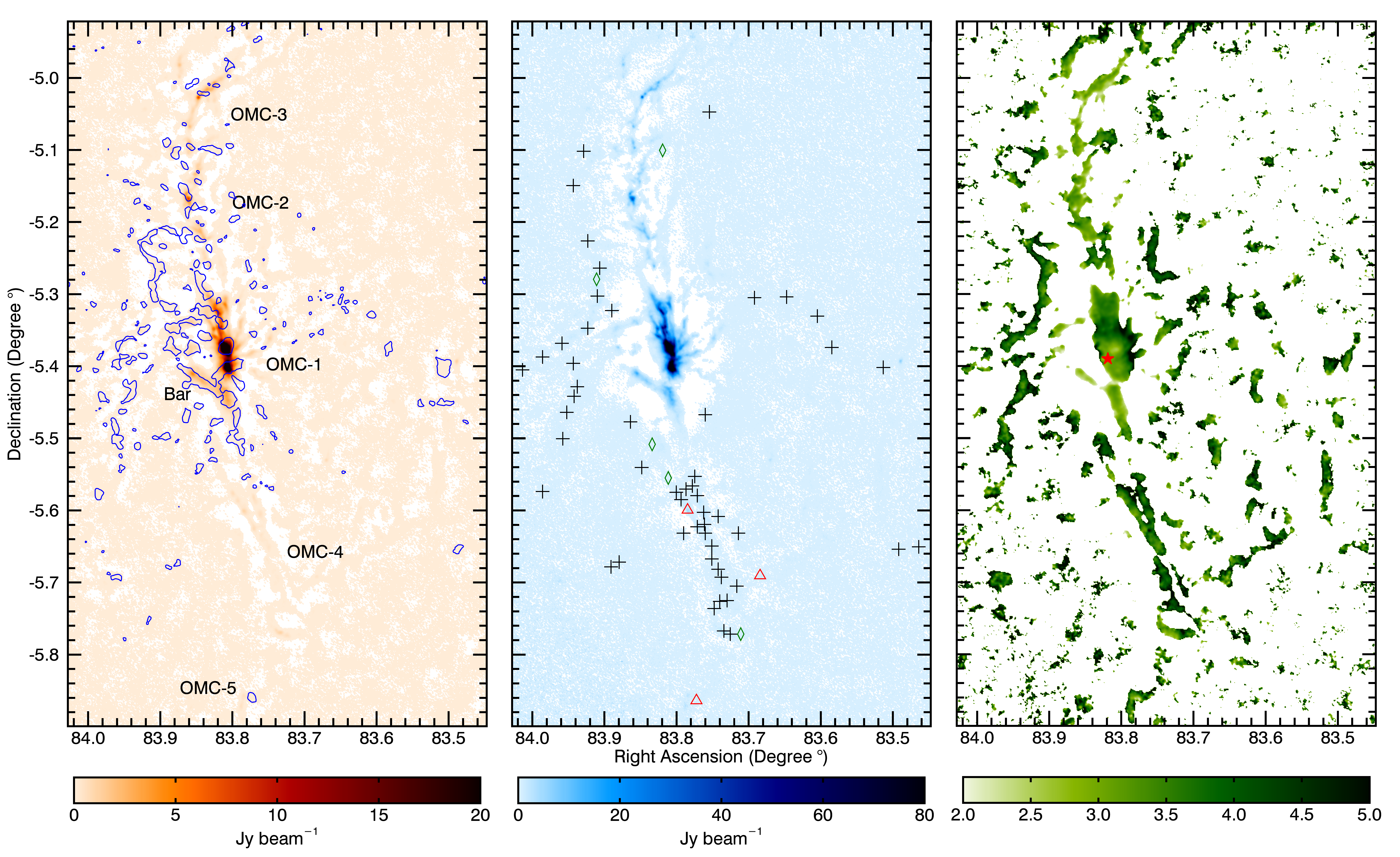}
\caption{The Orion A molecular cloud complex as seen by the SCUBA-2 shared-risk observations. \textit{Left}: Continuum map at 850 $\mu$m in Jy per unit of 850 $\mu$m beam area. The blue contour traces the 20 K km s$^{-1}$ emission from the $^{12}$CO J=3-2 molecular line. The labels identify the approximate location of different regions in the integral-shaped filament. \textit{Middle}: Continuum map at 450 $\mu$m map in Jy per unit of 450 $\mu$m beam area. The 60 submillimetre sources selected for this study (see Tables \ref{tab:01} and \ref{tab:02}) are identified as follows: clumps are black plus symbols, protostars are red triangles and young stars with disks are green diamonds. \textit{Right}: Map of the spectral index $\alpha$ as defined in Equation \ref{eq:02}. Its relationship to the spectral index of emissivity $\beta$ is given in Equation \ref{eq:03}. In the Rayleigh-Jeans approximation, this relation is simplified to $\beta_{RJ}=\alpha-2$. The red star identifies the position of $\theta$ Orionis C, the brightest member of the Trapezium cluster.}
\label{fig:02}
\end{minipage}
\end{figure*}

The Submillimetre Common-User Bolometer Array (SCUBA-2) is a 10,000 pixel cryogenically cooled continuum camera capable of simultaneously observing in the 450 and 850 $\mu$m atmospheric windows \citep{bib:12}. The Heterodyne Array Receiver Programme (HARP) is a 16-element detector providing high-sensitivity and high-resolution spectral line measurements between 325 and 375 GHz (799 and 922 $\mu$m). The Auto-Correlation Spectral Imaging System (ACSIS) is a digital interferometric spectrometer that can use HARP measurements to create large-scale velocity maps \citep{bib:13}.

Shared-risk observations of the Orion A molecular cloud complex were taken on February 13 2010 at the James Clerk Maxwell Telescope (JCMT) with SCUBA-2. The measured 225 GHz atmospheric opacity during the observations varied between $\tau_{225} = 0.036 - 0.045$. The complementary spectroscopic observations of the $^{12}$CO J=3-2 molecular line were obtained between February 9 and 16 in 2007 with HARP \citep{bib:32}. The opacity varied between $\tau_{225} = 0.079 - 0.094$. The three maps cover a field of approximately 60$\arcmin$ by 60$\arcmin$. They include most of the Orion A molecular clouds along the integral-shaped filament, such as OMC-1 to OMC-3, as well as OMC-4 and the northernmost part of OMC-5. 

The $^{12}$CO J=3-2 emission line map used in this study was reduced independently from the SCUBA-2 shared-risk observations \citep{bib:32}. The Orion A shared-risk observations were reduced using the Starlink software map-making capabilities \citep{bib:42,bib:43}. The shared-risk programme for SCUBA-2 was completed with only one array (instead of the final four) of the detector in each continuum band, leading to less sensitive maps than what can be produced by the full detector \citep[e.g.][Mairs et al. in preparation]{bib:36}. Because of differences in the way large-scale emission is handled between the shared-risk and the full-array reductions, the $^{12}$CO J=3-2 contamination analysis presented in this study applies only to small-scale structures ($< 1\arcmin$). Contamination levels for individual sources may change when larger spatial scales are included.

As mentioned in Section 2, some calibration is needed before the SCUBA-2 and HARP maps can be used together. Because of the way the telescope's dish distributes the incident energy over the detectors, the effective beam shape for each SCUBA-2 bandpass is approximated by two Gaussian components. The full-width at half-maximum (FHWM) of these main and secondary beams are, respectively, 7.9$\arcsec$ and 25$\arcsec$ at  450 $\mu$m  with relative amplitudes of 0.94 and 0.06. At 850 $\mu$m, they are 13.0$\arcsec$ and 48$\arcsec$  with relative amplitudes of 0.98 and 0.02 \citep{bib:39,bib:14}. The process of beam-matching consists of convolving each map with the other's effective beam. This is done to ensure that the 450 and 850 $\mu$m maps are comparable by degrading their quality in similar ways. The $^{12}$CO J=3-2 emission map is treated slightly differently. The map is first filtered with a Gaussian (FWHM = 1.0$\arcmin$) in order to subtract excess large-scale background \citep{bib:06}. This is made necessary by HARP's sensitivity to larger spatial scales than the 850 $\mu$m shared-risk observations. This large-scale emission can represent a significant proportion of the $^{12}$CO J=3-2 integrated intensity ($\approx 85 \%$ in some parts of the cloud). It is then convolved with the 450 $\mu$m primary and secondary beams to degrade it in a similar way to the 850 $\mu$m map. This was done to treat CO data in a manner as close as possible to the 850 $\mu$m data. Once the beam-matching process is complete, an automated linear interpolation transforms the 450 $\mu$m and $^{12}$CO J=3-2 maps to share the same pixel-scale (4.0$\arcsec$) and associated celestial coordinates as the 850 $\mu$m map. 

Before the three maps can be used with Equation \ref{eq:02}, they first need to be properly normalized. Since we are working with peak photometry (Jy beam$^{-1}$) instead of aperture photometry (Jy arcsec$^{-2}$), this calibration must be included in the beam-matching process. We use the effective beam widths $\theta_{E450} = 9.8\arcsec$ and $\theta_{E850} = 14.6\arcsec$ from \citet{bib:14} to calculate the new effective mixed beam width $\theta_{M} = 17.6\arcsec$ created by convolving the 450 and 850 $\mu$m beams together. After applying the appropriate filtering, each map is multiplied by a correction factor calculated from these effective beams to compensate for the change in beam area. Once this normalization is complete, the three maps effectively share the same units and can be directly compared.

The noise levels were calculated after the beam matching process, for an effective 17.6$\arcsec$ beam width, by measuring the standard deviation of the background emission in each map. This was done by selecting a 276$\arcsec$ by 172$\arcsec$ low-emission region centered at (5:34:19.550, -5:42:41.91). To calculate the spectral index map, only pixels with a flux density above 1$\sigma$ were used, which is 280 mJy beam$^{-1}$ at 450 $\mu$m and 20 mJy beam$^{-1}$ at 850 $\mu$m.

\section{Results}

\begin{figure}
\includegraphics[width=84mm]{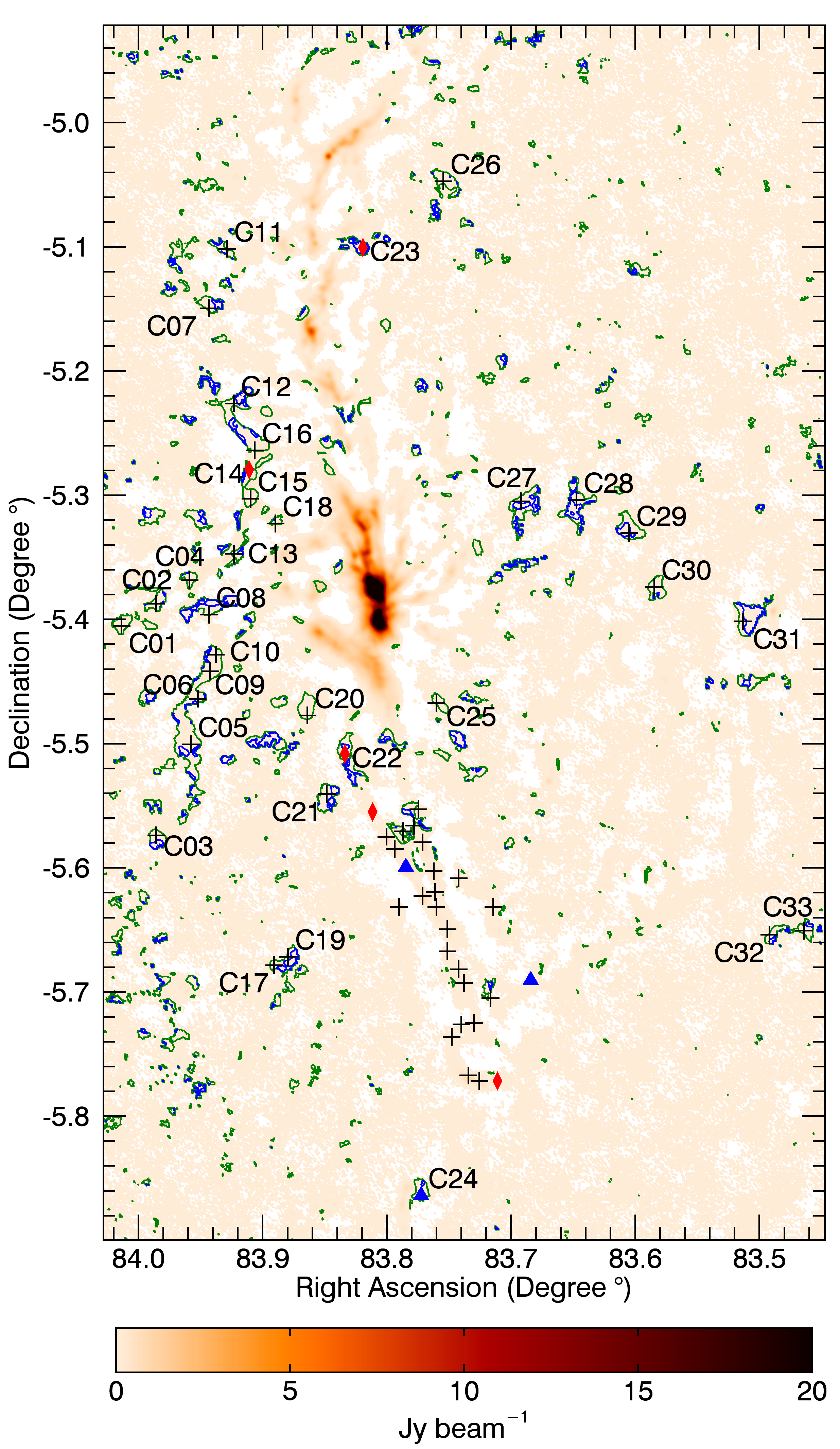}
\caption{Continuum map of the Orion A molecular cloud complex at 850 $\mu$m in Jy per unit of 850 $\mu$m beam area. The contours trace two levels of the $^{12}$CO J=3-2 molecular line contamination as a percentage of the corresponding 850 $\mu$m flux density (10$\%$ in green and 30$\%$ in blue). The 60 submillimetre sources selected for this study (see Tables \ref{tab:01} and \ref{tab:02}) are identified as follows: clumps are black plus symbols, protostars are blue triangles and young stars with disks are red diamonds. For ease of reading, only the locations of high $^{12}$CO contamination (i.e., Table \ref{tab:01}) are labelled.}
\label{fig:06}
\end{figure}

The shared-risk observations for Orion A are shown in Figure \ref{fig:02} for the 850 $\mu$m and 450 $\mu$m bands. The measured $^{12}$CO J=3-2 integrated intensity for a value of 20 K km s$^{-1}$ is shown as the blue contour in the left panel of Figure \ref{fig:02}. This figure also shows the emissivity spectral index $\alpha$ map from Equation \ref{eq:02} after the $^{12}$CO J=3-2 contamination was removed from the 850 $\mu$m map. An estimate of the regional variations in emissivity can be obtained from Equation \ref{eq:03} in the Rayleigh-Jeans approximation: $\beta_{RJ} = \alpha - 2$. As stated previously, this approximation generally leads to an incomplete representation of low temperature regions by underestimating the value of the spectral index $\beta$. However, it still shows some key features such as the low emissivity region in OMC-1 north-west of the Trapezium stars. This is a region with well-known contamination from molecular lines other than the $^{12}$CO J=3-2 line \citep{bib:52,bib:37,bib:38}. The Orion bar also shows low spectral index regions similar to those measured in previous studies \citep{bib:15,bib:16}.

The $^{12}$CO J=3-2 molecular line contamination is shown in Figure \ref{fig:06} overlaid on the 850 $\mu$m continuum map. The green contours trace three levels of the $^{12}$CO J=3-2 line emission as a percentage of the corresponding 850 $\mu$m flux density (10$\%$, 20$\%$ and 30$\%$). To study molecular contamination in more detail, we chose 33 sources around the integral-shaped filament selected by eye for their higher level of $^{12}$CO J=3-2 line emission. This represents most sources with a contamination $C_{850} \geq 0.1$ in Equation \ref{eq:02}. Each area of high contamination is clearly labelled in Figure \ref{fig:06}. The continuum properties of these sources were checked individually to verify their plausibility as clump candidates. We took care to select sources away from possibly problematic regions in the 450 and 850 $\mu$m maps. This is made necessary because the contrast between intensities in the SCUBA-2 maps can create bowling effects that artificially decrease the flux density around very bright regions (OMC-1, for example). We also use 27 clumps identified by \citet{bib:18} in OMC-4. As these authors suspected, we confirm that these clumps have very low levels of $^{12}$CO J=3-2 contamination and thus can be used as a control group. 

The positions of all 60 sources are indicated in the 450 $\mu$m continuum map of Figure \ref{fig:02}, as well as in Figure \ref{fig:06} (see caption). Their properties, shown in Tables \ref{tab:01} and \ref{tab:02}, are obtained by averaging the values in each small 12$\arcsec$ by 12$\arcsec$ map box centered on the peak position of the source. These boxes have roughly the same angular area as the effective 850 $\mu$m beam. Table \ref{tab:01} lists the 33 sources selected for their higher levels of $^{12}$CO J=3-2 molecular line contamination. Their IDs have been attributed in order of decreasing right ascension. Table \ref{tab:02} lists the 27 reference clumps in OMC-4. The ID for each clump is the same as given by \citet{bib:18}.

\begin{figure}
\includegraphics[width=84mm]{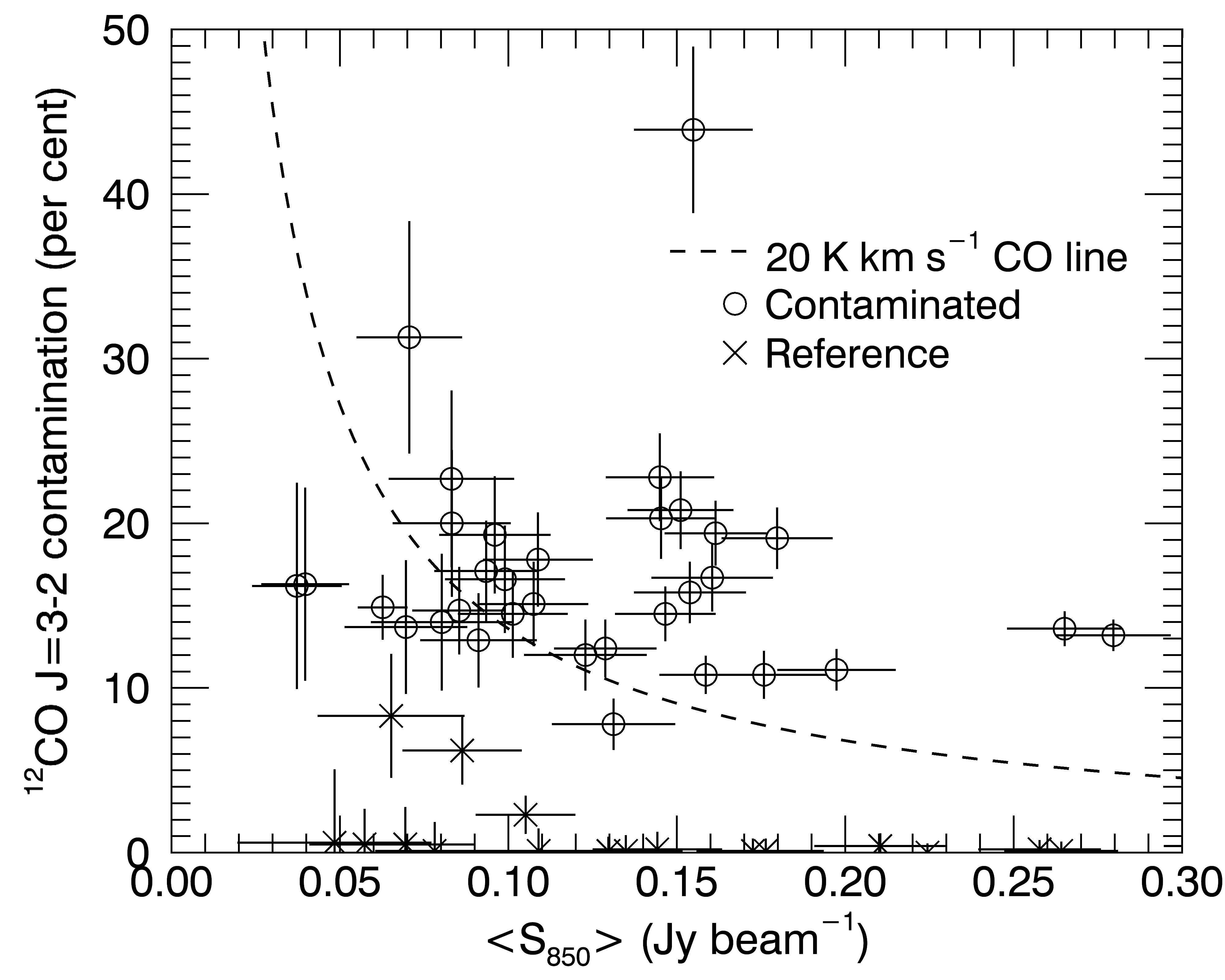}
\includegraphics[width=84mm]{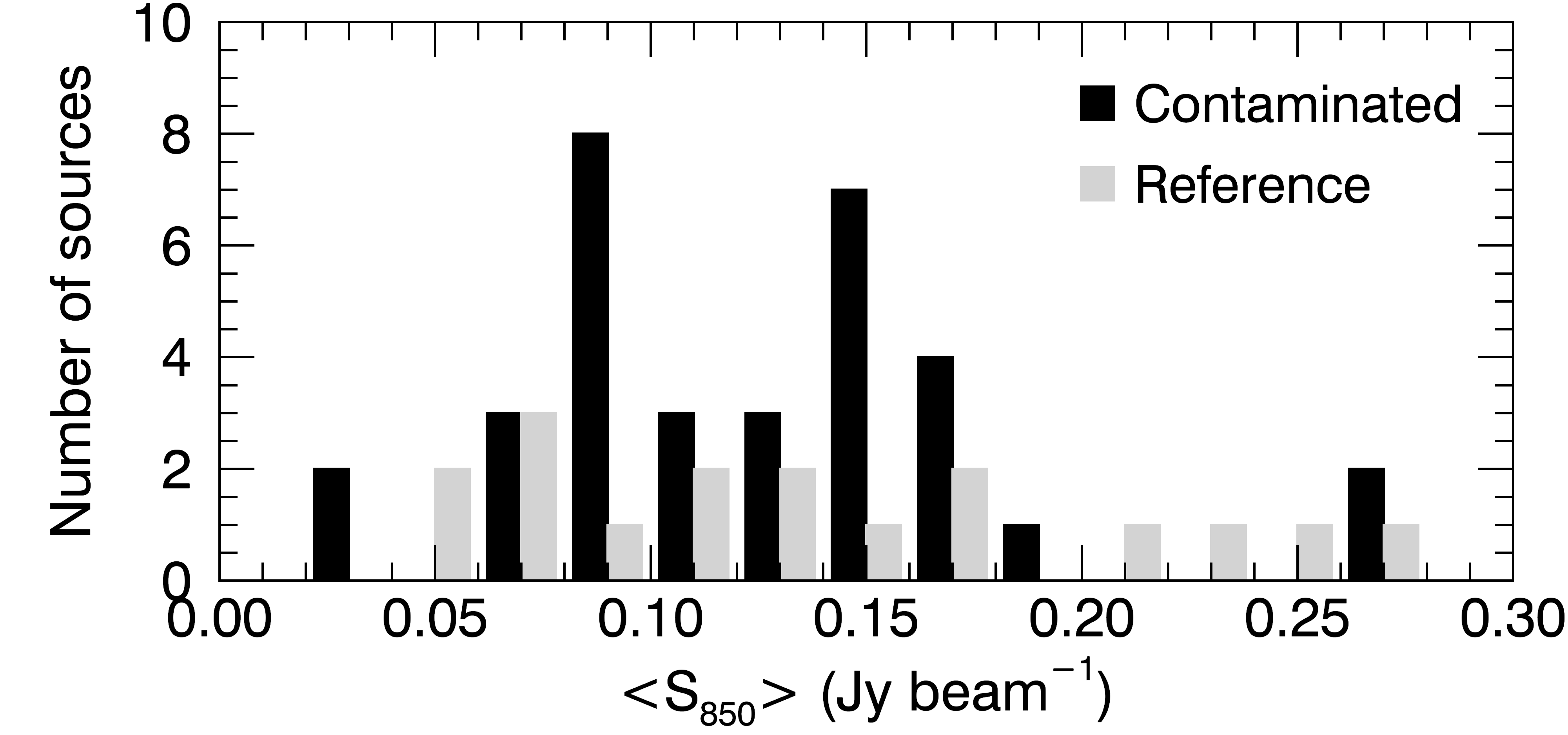}
\caption{\textit{Top}: $^{12}$CO J=3-2 molecular line emission as a percentage ($\%$) of the corresponding average flux density $\left\langle S_{850}\right\rangle$ (Jy beam$^{-1}$) measured directly from the SCUBA-2 850 $\mu$m shared-risk observation map. The dashed line sets the percentage of contamination for a $^{12}$CO J=3-2 line emission at 20 K km s$^{-1}$. The empty circles are the 33 sources chosen in this study for their high level of molecular contamination. The crosses are reference sources identified by \citet{bib:18} in OMC-4 with an average flux density lower than 0.3 Jy beam$^{-1}$. Since no reference source above 0.3 Jy beam$^{-1}$ shows significant contamination, the horizontal axis is truncated at that value to facilitate the viewing of this figure. The uncertainties for all sources are identified with plain black lines. \textit{Bottom}: Histogram showing the distribution of contaminated (black bars) and reference (gray bars) sources as a function of their average flux density $\left\langle S_{850}\right\rangle$ (Jy beam$^{-1}$). The categories have a width of 0.02 Jy beam$^{-1}$, with a range going from 0.00 to 0.30 Jy beam$^{-1}$.}
\label{fig:03}
\end{figure}

\begin{figure}
\includegraphics[width=84mm]{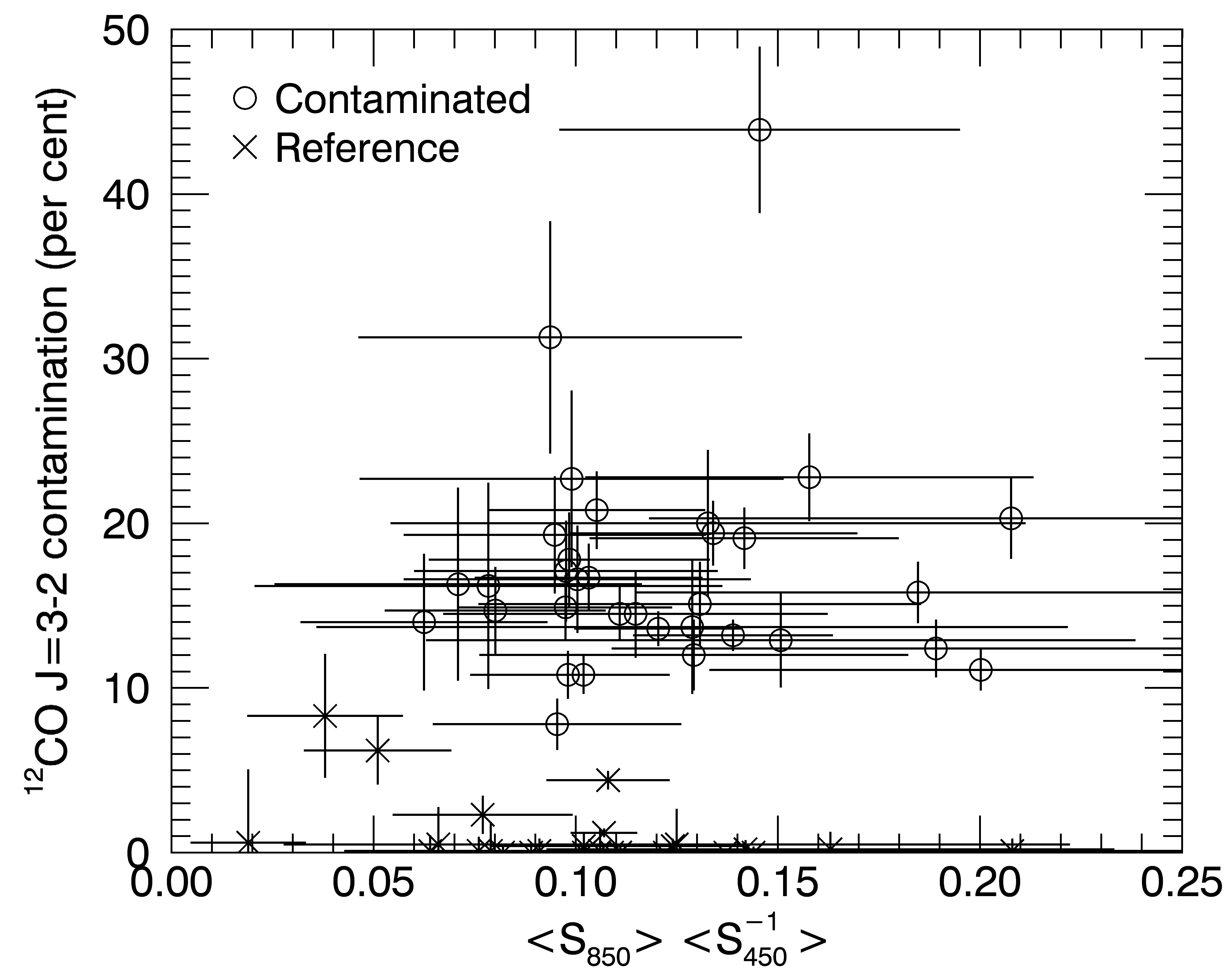}
\includegraphics[width=84mm]{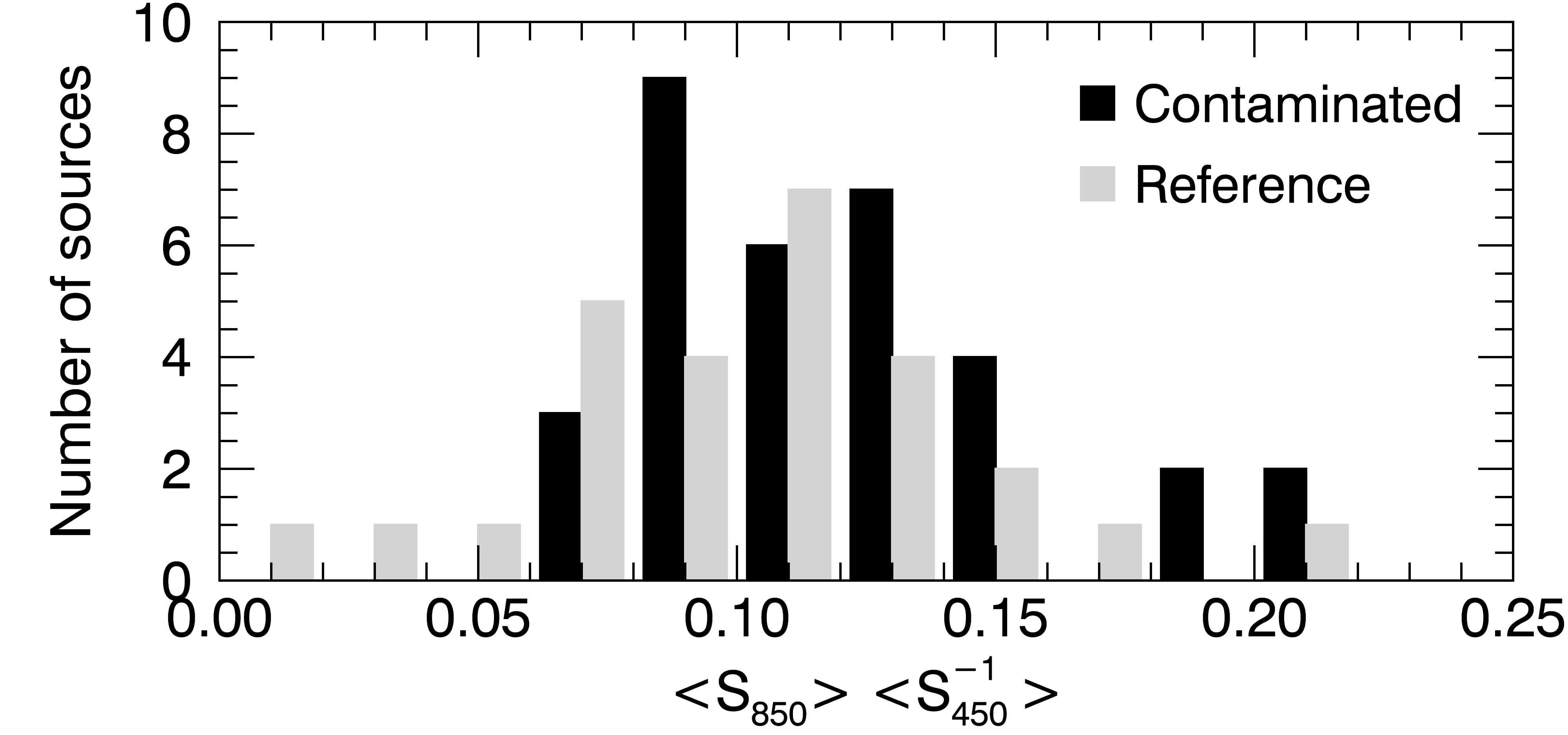}
\caption{\textit{Top}: $^{12}$CO J=3-2 molecular line emission as a percentage ($\%$) of the corresponding average 850 $\mu$m flux density $\left\langle S_{850}\right\rangle$, as a function of the measured (i.e., uncorrected for CO contamination) 850 to 450 $\mu$m average flux ratio $\left\langle S_{850}\right\rangle \left\langle S_{450}\right\rangle^{-1}$. The symbols have the same meaning as in Figure \ref{fig:03}. For each source, these two quantities are used in Equation \ref{eq:02} to deduce the spectral index $\alpha$ before and after contamination correction. The uncertainties for all sources are identified with plain black lines. \textit{Bottom}: Histogram showing the distribution of contaminated (black bars) and reference (gray bars) sources as a function of their average flux ratio $\left\langle S_{850}\right\rangle \left\langle S_{450}\right\rangle^{-1}$. The histogram bins have a width of 0.02, with a range going from 0.00 to 0.26.}
\label{fig:04}
\end{figure}

The $^{12}$CO J=3-2 contamination in each clump expressed as a percentage of their average 850 $\mu$m flux density is plotted versus their average 850 $\mu$m flux density per unit of effective 850 $\mu$m beam area ($\theta_{E850} = 14.6\arcsec$) in Figure \ref{fig:03}; the same contamination percentage is given in Figure \ref{fig:04} as a function of the 850 to 450 $\mu$m flux ratio. Although we use the beam matched ($\theta_{M} = 17.6\arcsec$) values to calculate the contamination and the ratio of flux densities (and thus the spectral indices), it is more useful for future studies to describe the sources presented in this paper by their $\left\langle S_{850}\right\rangle$ flux density per unit of effective 850 $\mu$m beam area. The $^{12}$CO J=3-2 contamination as a percentage of the 850 $\mu$m flux density remains essentially unchanged between the matched ($\theta_{M}$) and effective ($\theta_{E850}$) beams.

This list of 60 sources in Orion A was compared to a catalogue of known young stellar objects identified by \citet{bib:19} using the Spitzer Space Telescope. We confirm the presence of three protostars (blue triangles in Figure \ref{fig:06}) in our sample, including one (C24), showing a $^{12}$CO J=3-2 contamination of $16 \%$, in the northernmost part of OMC-5. Furthermore, five young stars with disks lie within continuum sources along the integral-shaped filament, including three in regions with strong $^{12}$CO J=3-2 line emission. The most highly contaminated source in this study (C23), with a level of $44 \%$, is associated with a young star with a disk located between OMC-2 and OMC-3 (identified in Figure \ref{fig:06}). It is unlikely that this object alone is responsible for such a strong $^{12}$CO J=3-2 line emission. Some possible explanations include that this region is a shock knot, or that it is surrounded by outflows from past or ongoing star formation. The objects in our sample that have also been detected with the Spitzer Space Telescope are identified on Figure \ref{fig:06} (see caption) and listed in Tables \ref{tab:01} and \ref{tab:02}. Additional notes on astronomical objects found near these sources of interest are presented in Appendix B.

To check for signs of molecular outflows, the wing criterion from \citet{bib:34} ($T_{MB}>1.5$ K at $3.0$ km s$^{-1}$ from line centre) was applied to the spectrum of each source from Tables \ref{tab:01} and \ref{tab:02}. This criterion is reached for most sources in both the contaminated and reference samples; the spectra often exhibit wide $^{12}$CO J=3-2 molecular lines. The wing criterion is not met in only a few cases: five sources in the contaminated sample (C12, C17, C19, C32, C33) and one in the reference sample (06). All the Spitzer-identified young stellar objects listed in Tables \ref{tab:01} and \ref{tab:02} also fit the wing criterion with wide $^{12}$CO J=3-2 lines. It is therefore likely that molecular outflows play a major role in $^{12}$CO J=3-2 line contamination at 850 $\mu$m.

Before we study the effect of molecular line contamination in more detail, we can deduce some general guidelines concerning $^{12}$CO J=3-2 emission in the sources shown on Figure \ref{fig:03}. First of all, all the identified sources with a contamination level above $10 \%$ seem to be concentrated below a threshold of 300 mJy beam$^{-1}$ in the 850 $\mu$m SCUBA-2 shared-risk map. Such a limit is to be expected because the optically thick $^{12}$CO J=3-2 line emission saturates much faster than the mostly optically thin dust thermal emission. The $^{12}$CO J=3-2 line emission is therefore more likely to become a problem toward regions of lower column density. This also explains why peaks in $^{12}$CO J=3-2 line emission do not necessarily lead to high levels of contamination in SCUBA-2 maps. The OMC-1 central $^{12}$CO J=3-2 emission peak is an example of this situation; it is completely dominated by the total 850 $\mu$m peak flux density: it produces a contamination $\leq 1\%$. This does not mean that the total molecular contamination in the central part of OMC-1 is negligible; it is a region known for its multitude of molecular species \citep{bib:37,bib:38} which may contribute up to $60\%$ of its 850 $\mu$m intensity \citep{bib:52}. There is however no easy way to guess which sources will be contaminated by molecular emission without having the corresponding spectroscopic observations. Even among sources with the same classification, as we see in Spitzer-identified protostars and young stars with disks in our sample, there can be wide differences in the levels of $^{12}$CO J=3-2 line emission. The sources that do show a strong molecular contamination must be corrected as their deduced physical properties depend on an accurate measurement of flux density.

The most direct effect of $^{12}$CO J=3-2 molecular line emission is its contribution to the total 850 $\mu$m flux density measured by SCUBA-2, therefore inflating the resulting 850 to 450 $\mu$m flux ratio. Following Equation \ref{eq:02} and Figure \ref{fig:01}, this systematically leads to an underestimation of the spectral indices $\alpha$ and $\beta$. The size of this deviation depends on the fraction of molecular line contamination in the measured 850 $\mu$m flux density, as shown in Equation \ref{eq:05}. The percentage of $^{12}$CO J=3-2 molecular line emission and the measured 850 to 450 $\mu$m flux ratios are shown in Figure \ref{fig:04} for a sample of 60 sources in Orion A. The largest spectral index difference ($\Delta\alpha=0.9^{+0.3}_{-0.1}$) is associated with an object (C23) with a $^{12}$CO J=3-2 line contamination percentage of $44\%$. Its average 850 to 450 $\mu$m flux ratio is reduced from $0.15\pm0.04$ to $0.08\pm0.03$, thus increasing the deduced spectral index $\alpha$ from $3.0^{+0.5}_{-0.4}$ to $3.9^{+0.6}_{-0.4}$.

\begin{figure}
\includegraphics[width=84mm]{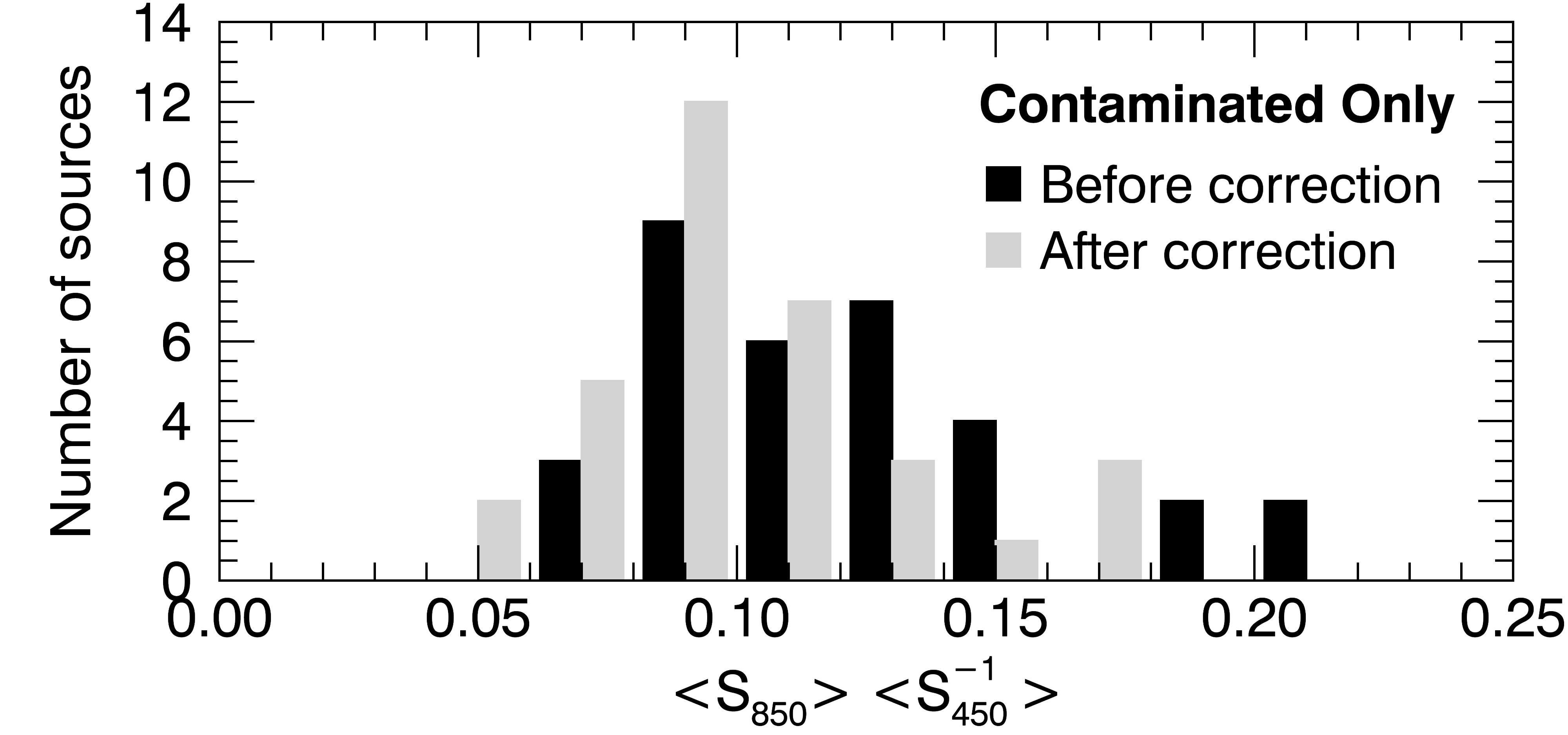}
\includegraphics[width=84mm]{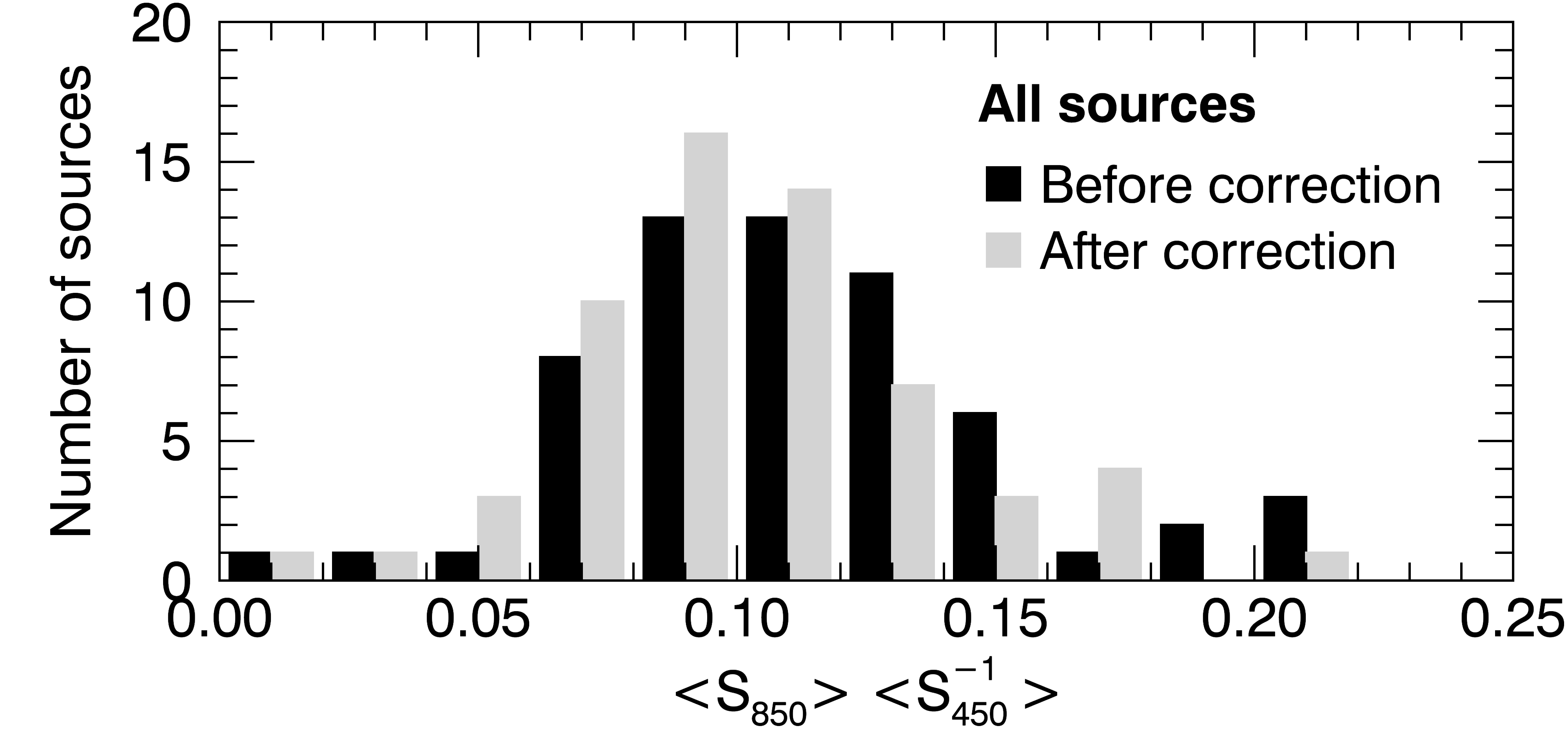}
\caption{Histograms showing the effect of contamination correction on the 850 to 450 $\mu$m average flux ratio $\left\langle S_{850}\right\rangle \left\langle S_{450}\right\rangle^{-1}$. The number of sources for each bin is given before (black bars) and after (gray bars) contamination correction is applied to the measured ratio. The histogram bins have a width of 0.02, with a range going from 0.00 to 0.26. \textit{Top}: The histogram is limited to the 33 sources chosen in this study for their high level of molecular contamination. \textit{Bottom}: The histogram including all 60 sources shown in Figure \ref{fig:04}.}
\label{fig:05}
\end{figure}

The effects of molecular contamination on emissivity can also be understood by looking at the histograms in Figure \ref{fig:05}. By using the flux ratio as a proxy for the spectral index $\beta$, following the same relation as the one shown in Figure \ref{fig:01}, we directly quantify the difference $^{12}$CO J=3-2 line subtraction has on the emissivity of both the total source selection and the contaminated subgroup.The top panel of Figure \ref{fig:05} quantifies how correcting the 850 $\mu$m flux density influences the 850 to 450 $\mu$m flux ratio for the 33 most highly contaminated sources in our sample. By combining Equation \ref{eq:02} with the measurements shown in Figure \ref{fig:04}, we calculate the spectral index $\alpha$ before and after contamination correction for all 60 sources. The results are presented in Tables \ref{tab:01} and \ref{tab:02}. The average uncorrected ratio of these sources leads to a spectral index $\left\langle\alpha\right\rangle = 3.3^{+0.6}_{-0.4}$. The corresponding average correction is $\left\langle\Delta\alpha\right\rangle = 0.3^{+0.2}_{-0.2}$. This results in an average corrected value $\left\langle\alpha\right\rangle = 3.6^{+0.6}_{-0.4}$ for the contaminated sources. In comparison, the spectral index for the OMC-4 control group gives a value $\left\langle\alpha_{O\!M\!C4}\right\rangle = 3.6^{+0.8}_{-0.5}$. The uncertainties are obtained from the standard deviations of the ratios. Although they are statistical uncertainties, we have just enough sources in each sample (33 and 27) for them to be significant. If these two samples are indeed comparable, as their similar standard deviations and their distribution of ratios in Figure \ref{fig:04} would suggest, then this indicates that the contamination correction brings the average spectral indices $\alpha$ to levels similar to those in the uncontaminated clumps of OMC-4. The similarity between the two samples could be explained by the selection of sources away from the ridge in OMC-1 (see Figure \ref{fig:02}), known to be warmer and denser than the rest of the integral shaped filament. 

A Kolmogorov-Smirnov test was applied to both the contaminated and the total subgroups from Figure \ref{fig:05} in order to verify the significance of the contamination correction on these distributions. The subgroup with the 33 contaminated sources has a $0.8 \%$ probability that the uncorrected and corrected distributions are statistically identical. Since it represents the group most affected by molecular contamination, this is a confirmation that contamination correction has a significant effect on the measured flux ratios. For the complete selection, this test gives a $6.4 \%$ probability. Unsurprisingly, this only means that, as more uncontaminated sources are added to the sample tested, the less statistically significant the average contamination becomes. 

\section{Discussion}

Quantifying the molecular line contamination in submillimetre observations, such as in SCUBA-2 850 $\mu$m maps, is a necessary step in the study of cold interstellar dust. Through the construction of accurate spectral energy distributions, it is possible to learn much about the physical properties of this important mass tracer in giant molecular clouds. As an example, models using observations carried out with the Herschel Space Observatory, hereafter Herschel, benefit strongly from the addition of SCUBA-2 850 $\mu$m measurements. While Herschel observations efficiently probe the peak blackbody emission of cold dust grains, SCUBA-2 wavelengths help constrain the Rayleigh-Jeans tail of the distribution. This leads to a better determination of density, temperature and spectral index $\beta$ than Herschel results alone \citep{bib:28}. Furthermore, both the JCMT and Herschel have significant overlapping data in the Gould Belt.

A good example of what can be achieved with a multifaceted approach is the work of \citet{bib:54}. These authors combined observations from the Planck and Herschel space telescopes in order to build optical depth and temperature maps for the Orion molecular cloud complex. They also included in their fits the spectral index $\beta$ map from \citet{bib:58} at a resolution of $36 \arcmin$. From this optical depth map, and with near-infrared extinction measurements from 2MASS, \citet{bib:54} obtained a column density map for the region. Finally, they found a temperature $T_d \approx 15$ K in dense regions along the spine of Orion A, which we will use as a reference for the examples used in this discussion.

Since dust thermal emission depends strongly on the emissivity of the grains, we focus specifically on how molecular contamination influences this property. This is why we use the spectral index $\beta$ instead of $\alpha$ for the rest of this discussion. We have already given in Section 2 a relation between the spectral indices $\alpha$ and $\beta_{RJ}$ obtained from Equation \ref{eq:03} in the Rayleigh-Jeans approximation: $\beta_{RJ} = \alpha-2$. This allows us to use these indices interchangeably to calculate differences ($\Delta\alpha = \Delta\beta$), assuming the dust temperature $T_d$ is fixed. The Rayleigh-Jeans approximation also provides us with a reliable lower limit for the more realistic emissivity spectral index $\beta$.

\begin{table*}
\begin{minipage}{145mm}
\caption{Measured properties of 33 sources selected for their $^{12}$CO J=3-2 contamination}

\begin{tabular}{| c | c | c | c | c | c | c | c | c | c | c | c | c |}

\hline
ID & RA$^1$ & Dec$^1$ & $\left\langle S_{850} \right\rangle$$^2$ & $\sigma_{850}$$^2$ & $\%$$^3$ & $\sigma_{\%}$$^3$ & Ratio$^4$ & $\sigma_R$$^4$ & $\alpha$$^5$ & $\left\langle \sigma_{\alpha} \right\rangle$$^5$ & $\Delta\alpha$$^6$ & YSO$^7$\\
\hline
C01&	05 36 03.24&	-05 24 18.0&	108&	16&	15&	3&	0.11&	0.04&	3.5&	0.6&	0.3\\
C02&	05 35 56.54&	-05 23 14.0&	101&	16&	15&	3&	0.10&	0.04&	3.7&	0.6&	0.2\\
C03&	05 35 56.55&	-05 34 26.0&	145&	16&	20&	3&	0.17&	0.06&	2.8&	0.6&	0.4\\
C04&	05 35 50.11&	-05 22 06.0&	147&	15&	14&	2&	0.09&	0.02&	3.7&	0.4&	0.2\\
C05&	05 35 49.85&	-05 30 02.0&	180&	16&	19&	2&	0.11&	0.03&	3.4&	0.4&	0.3\\
C06&	05 35 48.51&	-05 27 50.0&	160&	18&	17&	3&	0.09&	0.02&	3.9&	0.4&	0.3\\
C07&	05 35 46.35&	-05 08 58.0&	70&	18&	14&	5&	0.11&	0.07&	3.5&	1.1&	0.2\\
C08&	05 35 46.36&	-05 23 46.0&	176&	18&	11&	2&	0.09&	0.02&	3.8&	0.4&	0.2\\
C09&	05 35 46.10&	-05 26 30.0&	265&	17&	14&	1&	0.10&	0.02&	3.6&	0.3&	0.2\\
C10&	05 35 45.02&	-05 25 42.0&	280&	17&	13&	1&	0.12&	0.02&	3.3&	0.3&	0.2\\
C11&	05 35 42.87&	-05 06 06.0&	131&	18&	8&	2&	0.09&	0.03&	3.8&	0.4&	0.1\\
C12&	05 35 41.53&	-05 13 34.0&	96&	16&	19&	4&	0.08&	0.03&	4.0&	0.6&	0.3\\
C13&	05 35 41.54&	-05 20 50.0&	63&	7&	15&	2&	0.08&	0.02&	3.9&	0.4&	0.3\\
C14&	05 35 38.59&	-05 16 46.1&	85&	14&	15&	3&	0.07&	0.02&	4.2&	0.5&	0.2& YSD\\
C15&	05 35 38.32&	-05 18 10.1&	161&	15&	19&	2&	0.11&	0.03&	3.5&	0.4&	0.3\\
C16&	05 35 37.52&	-05 15 50.1&	159&	14&	11&	2&	0.09&	0.02&	3.8&	0.3&	0.2\\
C17&	05 35 33.78&	-05 40 42.0&	83&	17&	20&	5&	0.11&	0.06&	3.5&	0.9&	0.4\\
C18&	05 35 33.50&	-05 19 22.1&	197&	17&	11&	2&	0.18&	0.05&	2.7&	0.5&	0.2\\
C19&	05 35 31.10&	-05 40 18.1&	80&	21&	14&	5&	0.05&	0.03&	4.6&	0.7&	0.2\\
C20&	05 35 27.35&	-05 28 38.1&	93&	15&	17&	4&	0.08&	0.03&	4.0&	0.5&	0.3\\
C21&	05 35 23.60&	-05 32 26.1&	109&	16&	18&	3&	0.08&	0.03&	4.0&	0.5&	0.3\\
C22&	05 35 20.11&	-05 30 30.1&	40&	13&	16&	6&	0.06&	0.04&	4.4&	1.0&	0.3& YSD\\
C23&	05 35 16.63&	-05 06 02.1&	155&	17&	44&	5&	0.08&	0.03&	3.9&	0.6&	0.9& YSD\\
C24&	05 35 05.37&	-05 51 50.0&	154&	16&	16&	2&	0.16&	0.05&	2.9&	0.5&	0.3& P\\
C25&	05 35 02.43&	-05 28 02.1&	129&	15&	12&	2&	0.17&	0.06&	2.8&	0.6&	0.2\\
C26&	05 35 01.10&	-05 02 50.1&	37&	13&	16&	7&	0.07&	0.04&	4.3&	1.1&	0.3\\
C27&	05 34 46.10&	-05 18 18.0&	83&	18&	23&	6&	0.08&	0.04&	4.0&	0.8&	0.4\\
C28&	05 34 35.39&	-05 18 14.0&	99&	18&	17&	4&	0.08&	0.03&	3.9&	0.6&	0.3\\
C29&	05 34 25.21&	-05 19 50.0&	71&	15&	31&	8&	0.06&	0.03&	4.3&	0.8&	0.6\\
C30&	05 34 20.38&	-05 22 25.9&	145&	16&	23&	3&	0.12&	0.04&	3.3&	0.5&	0.4\\
C31&	05 34 03.24&	-05 24 05.8&	151&	15&	21&	3&	0.08&	0.02&	3.9&	0.4&	0.4\\
C32&	05 33 58.12&	-05 39 13.8&	123&	18&	12&	3&	0.11&	0.04&	3.4&	0.6&	0.2\\
C33&	05 33 51.42&	-05 39 01.7&	91&	17&	13&	3&	0.13&	0.06&	3.2&	0.8&	0.2\\
\hline

\end{tabular}

$^1$ Right Ascension and Declination of the sources for epoch J$2000.0$\\ 
$^2$ Average flux density and uncertainty (in mJy) per unit of effective 850 $\mu$m beam area (beam width of $14.6\arcsec$)\\
$^3$ $^{12}$CO J=3-2 molecular emission and uncertainty as a percentage of the 850 $\mu$m flux density\\
$^4$ Ratio of 850 to 450 $\mu$m flux densities corrected for $^{12}$CO J=3-2 contamination and uncertainty\\
$^5$ Spectral index corrected for $^{12}$CO J=3-2 contamination as shown in Equation \ref{eq:02} and average uncertainty\\
$^6$ Spectral index difference after correction for $^{12}$CO J=3-2 line emission\\
$^7$ Young Stellar Object: Protostar (P) or Young Star with Disk (YSD)\\
\label{tab:01}
\end{minipage}
\end{table*}

First, the results given in Section 4 need to be put into proper context. The dust in the diffuse interstellar medium is generally expected to have an emissivity spectral index near $\beta=1.7$ \citep{bib:24}. This has been confirmed by the \citet{bib:22} paper, where they measured a mean value of $\beta=1.62 \pm 0.10$ at high galactic latitudes. In molecular clouds and their dense filaments, the emissivity is however usually assumed constant at $\beta=2.0$ \citep{bib:20,bib:21, bib:36}. The spectral index in collapsing cores can vary between $\beta = 1.7 - 2.0$, while in protostellar disks it drops quickly below $\beta \leq 1.7$ as the grain sizes increase \citep{bib:10}. The emissivity can also change depending on the region studied. As an example, there is strong evidence of larger dust grains decreasing the emissivity in some prestellar cores of OMC-2 and OMC-3 \citep{bib:29}. The presence of ice mantles on dust grains can also, in theory, play a crucial role for the emissivity as they tend to increase the spectral index $\beta$ \citep{bib:08,bib:09}.

\begin{table*}
\begin{minipage}{145mm}
\caption{Measured properties of 27 clumps identified in OMC-4 by \citet{bib:18}}

\begin{tabular}{| c | c | c | c | c | c | c | c | c | c | c | c | c |}

\hline
ID & RA$^1$ & Dec$^1$ & $\left\langle S_{850} \right\rangle$$^2$ & $\sigma_{850}$$^2$ & $\%$$^3$ & $\sigma_{\%}$$^3$ & Ratio$^4$ & $\sigma_R$$^4$ & $\alpha$$^5$ & $\left\langle \sigma_{\alpha} \right\rangle$$^5$ & $\Delta\alpha$$^6$ & YSO$^7$\\
\hline
01&	05 34 44.21&	-05 41 26.0&	258&	18&	0&	1&	0.14&	0.03&	3.1&	0.3&	0.0& P\\
02&	05 34 50.63&	-05 46 18.0&	105&	14&	2&	2&	0.07&	0.02&	4.1&	0.4&	0.0& YSD\\
03&	05 34 51.44&	-05 37 54.0&	57&	16&	0&	3&	0.12&	0.08&	3.3&	1.1&	0.0\\
04&	05 34 51.98&	-05 42 18.0&	69&	20&	1&	3&	0.07&	0.03&	4.3&	0.7&	0.0\\
05&	05 34 54.12&	-05 46 18.0&	438&	15&	0&	1&	0.14&	0.02&	3.0&	0.2&	0.0\\
06&	05 34 55.19&	-05 43 30.0&	135&	16&	0&	1&	0.06&	0.02&	4.3&	0.3&	0.0\\
07&	05 34 56.26&	-05 46 02.0&	634&	15&	0&	1&	0.14&	0.01&	3.1&	0.1&	0.0\\
08&	05 34 57.07&	-05 41 34.1&	224&	16&	0&	1&	0.08&	0.01&	3.9&	0.2&	0.0\\
09&	05 34 57.60&	-05 43 34.0&	48&	29&	1&	5&	0.02&	0.02&	6.2&	1.3&	0.0\\
10&	05 34 58.14&	-05 36 30.1&	78&	17&	0&	2&	0.08&	0.03&	4.0&	0.6&	0.0\\
11&	05 34 58.14&	-05 40 54.1&	173&	16&	0&	1&	0.09&	0.02&	3.8&	0.3&	0.0\\
12&	05 34 59.48&	-05 44 10.1&	109&	19&	0&	2&	0.11&	0.04&	3.5&	0.6&	0.0\\
13&	05 35 00.29&	-05 38 58.1&	311&	16&	0&	1&	0.12&	0.02&	3.3&	0.2&	0.0\\
14&	05 35 00.29&	-05 40 02.1&	210&	19&	0&	1&	0.10&	0.02&	3.6&	0.3&	0.0\\
16&	05 35 02.70&	-05 37 10.1&	130&	14&	0&	1&	0.08&	0.02&	4.1&	0.3&	0.0\\
17&	05 35 02.43&	-05 37 54.1&	264&	17&	0&	1&	0.09&	0.01&	3.8&	0.2&	0.0\\
18&	05 35 02.97&	-05 36 10.1&	478&	16&	0&	1&	0.10&	0.01&	3.6&	0.1&	0.0\\
19&	05 35 05.11&	-05 37 22.1&	513&	17&	0&	1&	0.11&	0.01&	3.5&	0.1&	0.0\\
20&	05 35 05.11&	-05 34 46.1&	453&	18&	0&	1&	0.09&	0.01&	3.8&	0.2&	0.0\\
21&	05 35 05.91&	-05 33 10.1&	336&	18&	4&	1&	0.10&	0.02&	3.6&	0.2&	0.1\\
22&	05 35 06.72&	-05 33 58.1&	86&	17&	6&	2&	0.05&	0.02&	4.8&	0.5&	0.1\\
24&	05 35 08.33&	-05 35 58.1&	835&	18&	0&	1&	0.12&	0.01&	3.3&	0.1&	0.0& P\\
28&	05 35 08.86&	-05 34 14.1&	65&	22&	8&	4&	0.04&	0.02&	5.3&	0.8&	0.1\\
32&	05 35 09.66&	-05 37 54.1&	176&	17&	0&	1&	0.21&	0.07&	2.5&	0.5&	0.0\\
34&	05 35 10.47&	-05 35 06.1&	466&	17&	0&	1&	0.11&	0.01&	3.5&	0.2&	0.0\\
36&	05 35 12.08&	-05 34 30.1&	560&	18&	1&	1&	0.11&	0.01&	3.5&	0.1&	0.0\\
39&	05 35 14.76&	-05 33 18.1&	144&	19&	0&  1&	0.16&	0.06&	2.9&	0.6&	0.0& YSD\\
\hline

\end{tabular}

$^1$ Right Ascension and Declination of the sources for epoch J$2000.0$\\ 
$^2$ Average flux density and uncertainty (in mJy) per unit of effective 850 $\mu$m beam area (beam width of $14.6\arcsec$)\\
$^3$ $^{12}$CO J=3-2 molecular emission and uncertainty as a percentage of the 850 $\mu$m flux density\\
$^4$ Ratio of 850 to 450 $\mu$m flux densities corrected for $^{12}$CO J=3-2 contamination and uncertainty\\
$^5$ Spectral index corrected for $^{12}$CO J=3-2 contamination as shown in Equation \ref{eq:02} and average uncertainty\\
$^6$ Spectral index difference after correction for $^{12}$CO J=3-2 line emission\\
$^7$ Young Stellar Object: Protostar (P) or Young Star with Disk (YSD)\\

\label{tab:02}
\end{minipage}
\end{table*}

From this perspective alone, the average Rayleigh-Jeans emissivity spectral index $\beta_{RJ}$ obtained in Section 4 after correction for $^{12}$CO J=3-2 contamination ($\left\langle \beta_{RJ} \right\rangle = 1.6^{+0.6}_{-0.4}$) could be interpreted as at the lower limit for cores and filaments. While our spectral index map in Figure \ref{fig:02} is very similar to the one presented by \citet{bib:16}, the mean corrected indices for the contaminated and control samples ($\beta_{R\!J}\approx1.6$) are in fact slightly higher than expected from using the temperature-independent Rayleigh-Jeans approximation. Even if this approximation is useful to establish lower limits, we showed in Figure \ref{fig:01} that it underestimates the spectral index toward cold regions. As an example, a prestellar core with a real emissivity spectral index $\beta=2.0$ at a temperature $T_d = 15$ K would have, using Equation \ref{eq:03}, an associated spectral index $\alpha=3.0$ ($\beta_{R\!J}=1.0$). This illustrates the importance of knowing the temperature for finding a reliable value of the emissivity spectral index $\beta$. 

A possible explanation for the statistical uncertainties in the average measured Rayleigh-Jeans emissivity spectral index $\left\langle \beta_{RJ} \right\rangle = 1.6^{+0.6}_{-0.4}$ is that we are comparing objects of different natures. Since the properties of most sources in our sample are unknown, they could very well cover a large intrinsic range of emissivity spectral indices $\beta$ as well as dust temperatures $T_d$. This spread of measured spectral indices could also be explained in part by the uncertainties on the measured ratios, as they are shown in Figure \ref{fig:04}. Since the ratio method presented in this work is sensitive to small flux variations, these uncertainties are mostly driven by the noise level in the 450 $\mu$m flux density map. Furthermore, if the 450 $\mu$m map is more susceptible to large-scale fluctuations than its 850 $\mu$m counterpart, it could contribute to the artificial decrease in the measured ratios, thus increasing the average measured spectral indices for each sample. Although this leads to a large range of values for the Rayleigh-Jeans emissivity spectral index $\beta_{RJ}$, we know from Equations \ref{eq:05} and \ref{eq:03} that the correction $\Delta\beta$ for each source depends on the more accurate HARP measurements and on the smoother 850 $\mu$m map. We are confident that this measured effect of molecular contamination on the emissivity spectral index $\beta$ is a reliable estimate, even if the spectral indices themselves are subject to large uncertainties and probably also intrinsic spread.

The variation in the average spectral index due to $^{12}$CO J=3-2 line contamination correction $\left\langle\Delta\beta\right\rangle = 0.3^{+0.2}_{-0.2}$ illustrates the different scenarios possible in sources contaminated by molecular line emission. For some, it only acts as a correction leading to a slightly more accurate emissivity spectral index $\beta$. There are however cases where the difference is dramatic enough to change the physical description of the dust mixture in a submillimetre source. As an example, the most highly contaminated source in the sample (C23; $44\%$), associated with a Spitzer-identified young star with disk, gives a large correction $\Delta\beta=0.9_{-0.1}^{+0.3}$ that brings it on par with the average Rayleigh-Jeans spectral index measured in OMC-4 ($\beta_{R\!J}\approx1.6$). This $\Delta\beta \approx 0.9$ could mean the difference between grain growth up to centimetre sizes typical of protostellar disks ($\beta < 1.0$), and the emissivity spectral index expected in collapsing cores ($\beta > 1.7$) \citep{bib:10}. The other contaminated Spitzer objects (C14, C22, C24; $16 \pm 1 \%$) all share a correction $\Delta\beta\approx0.3$ comparable to the average difference $\left\langle\Delta\beta\right\rangle$ obtained from the entire contaminated sample. As discussed in the previous section, objects of these types (protostars and young stars with disks) are likely to be associated with molecular outflows, which would explain their $^{12}$CO J=3-2 line contamination levels \citep{bib:34}.

The dust temperature $T_d$ was assumed fixed up to this point for the analysis of the sources presented in this paper. If we instead suppose a fixed emissivity spectral index $\beta=2.0$ \citep{bib:44, bib:45, bib:36}, it is possible to rewrite Equations \ref{eq:05} and \ref{eq:03} to evaluate numerically the effect of molecular line contamination on the temperature determination ($\Delta T_d$). For the previous example of a theoretical submillimetre source with an emissivity spectral index $\beta=2.0$ and a temperature $T_d = 15$ K, a contamination level of $\approx 15\%$ ($\Delta \alpha = 0.3$) would then lead to a measured temperature $T = 12$ K. This temperature determination is however sensitive to the slightest variation in the 850 to 450 $\mu$m flux ratio. It becomes unreliable as the temperature increases and the thermal emission approaches the Rayleigh-Jeans regime. 

While the 850 to 450 $\mu$m flux ratio method is relevant to evaluate the effects of molecular contamination on the emissivity spectral index $\beta$, it is insufficient to properly derive the physical properties of cold interstellar dust grains. It requires some prior assumptions about the emissivity or the temperature, and it is insensitive to line-of-sight density variations \citep{bib:26}. A more accurate determination of those properties requires a multi-wavelength approach \citep{bib:27,bib:17}. Even then, molecular line contamination can remain a concern in spectral energy distributions built using SCUBA-2 850 $\mu$m observations. As an example, the dust temperature $T_d$ in cold interstellar regions can be well constrained by fitting the peak thermal emission with Herschel observations alone. On the other hand, the determination of the emissivity spectral index $\beta$ benefits from the addition of accurate SCUBA-2 850 $\mu$m measurements. In their study of the B1 clump in the Perseus molecular cloud, \citet{bib:28} showed that combining observations from both Herschel and SCUBA-2 lead to a better fit than using the Herschel data alone for the spectral energy distribution.

The degeneracy between the spectral index $\beta$ and the temperature $T_d$ is a serious challenge when attempting to fit accurate spectral energy distributions. \citet{bib:57} have shown that noise from observations can induce an artificial anti-correlation between $\beta$ and $T_d$ when using a $\chi^2$ fitting method. This is confirmed by \citet{bib:55}, who then introduced a hierarchical Bayesian fitting technique as an alternative to the $\chi^2$ method. The hierarchical Bayesian technique recovers the parameters of an artificial distribution with more accuracy; it is less sensitive to statistical noise and calibration uncertainties. It has also been successfully tested on Herschel observations of CB244. Furthermore, \citet{bib:56} compared several fitting methods and found they all exhibited some level of bias when estimating the relationship between $\beta$ and $T_d$. In particular, the hierarchical Bayesian model presented in \citet{bib:55}, while more precise than the $\chi^2$ method, is nonetheless biased toward a flat $\beta(T_d)$ relation in sources with a low signal-to-noise ratio.

It is important to point out other possible sources of contamination in the SCUBA-2 observations. Although generally much fainter than $^{12}$CO lines, the contribution from other molecular lines, such as those from $^{13}$CO and C$^{18}$O \citep{bib:35,bib:41, bib:51}, can add up to a significant portion of the total flux at both 450 and 850 $\mu$m toward hot cores. The Orion KL hot core in OMC-1 is famous for having a forest of molecular lines \citep{bib:37, bib:38} contributing a significant fraction ($\approx 60\%$) of its total flux density at 850 $\mu$m \citep{bib:52}. This is also true at 450 $\mu$m, where the corresponding molecular lines contribute up to $\approx 15\%$ of the total flux density toward Orion KL \citep{bib:53}.

Another common source of contamination is the continuum free-free emission from electrons in the interstellar medium. While it generally represents only a small fraction ($\leq1\%$) of the submillimetre emission at 850 $\mu$m in the galactic plane \citep{bib:23}, it can become non-negligible toward regions with strong photo-ionization. As an example, free-free emission in extended HII optically thin regions typically follow a gentle, almost flat, power-law \citep{bib:48, bib:49}. It is therefore likely to be measurable even at submillimetre wavelengths. However, this is not necessarily true for partially-thick jets or stellar winds where the free-free spectrum can be steeper \citep{bib:50}. In those cases, the contamination will depend on the properties of the observed object. One such example is the MWC 297 Herbig B1.5 Star in the Serpens molecular cloud, which exhibits free-free contamination of $73\%$ and $82\%$ at 450 and 850 $\mu$m respectively \citep{bib:45}.

Several regions in the Orion A molecular cloud show a strong free-free continuum. \citet{bib:25} fitted the dust thermal emission and the free-free emission toward three locations in OMC-1. The highest free-free contamination they found at 850 $\mu$m ($\approx 10 \%$)  is toward the free-free emission peak near the Trapezium stars. For Orion KL and Orion South, the free-free emission has only a small contribution ($\le 5\%$ of the total flux density at 850 $\mu$m). In each case, the free-free component at 450 $\mu$m is negligible ($<1\%$) when compared to the dust thermal emission. \citet{bib:15} also estimated free-free contamination in OMC-1 observations at 1110 $\mu$m. This contamination reaches up to $75\%$ in the HII region south-east of the Trapezium stars, and $\approx10 \%$ in the Orion bar. If we use these results as upper limits for the free-free contamination at 850 $\mu$m, it is possible to estimate its effect on the spectral index $\beta$ that would be calculated from SCUBA-2 observations. If free-free emission is negligible at 450 $\mu$m, we know from Equation \ref{eq:05} that contamination levels of $\approx 75\%$ and $\approx10\%$ at 850 $\mu$m would overestimate the spectral index by $\Delta\beta \approx 2.2 $ and $\Delta\beta \approx 0.2$ respectively. However, considering the strong free-free emission toward the HII region, it is possible given a flat power-law that this contribution is not negligible at 450 $\mu$m after all. If the 450 $\mu$m emission has some component of free-free emission, this would reduce the bias in the measurement of beta, but would not eliminate it completely.

\section{Conclusion}

We have presented the SCUBA-2 shared-risk observations for the Orion A molecular cloud complex at 450 and 850 $\mu$m. They were combined with HARP spectroscopic measurements in order to evaluate the effects of molecular contamination on the emissivity spectral index $\beta$. We studied a list of 33 sources along the integral-shaped filament chosen for their high $^{12}$CO J=3-2 line emission. At least four young stellar objects identified with the Spitzer Space Telescope [three young stars with disks (C14, C22, C23) and one protostar (C24)] show significant levels of contamination in their measured 850 $\mu$m flux densities. From the analysis of these contaminated sources, we have concluded that $^{12}$CO J=3-2 line contamination leads to an average underestimation $\left\langle\Delta\beta\right\rangle = 0.3^{+0.2}_{-0.2}$ in affected sources, with some individual sources displaying larger differences (up to $\Delta\beta=0.9_{-0.1}^{+0.3}$). While there is no obvious continuum property to easily identify contaminated sources, these results illustrate the need to subtract $^{12}$CO J=3-2 line emission from SCUBA-2 850 $\mu$m continuum maps where the corresponding HARP measurements are available. The properties shared by the contaminated sources seem to be their peak emission threshold ($S_{850} \le 300$ mJy beam$^{-1}$) and their more likely location toward regions of low dust column density. Emissivity is a crucial parameter for the description of dust grains in the interstellar medium, as it is related directly to size distributions in dust mixtures. Measuring it accurately is a necessary step in order to reach, with far-infrared and submillimetre observations, a more complete understanding of the physical processes leading to the formation of stars and their planetary systems.

Special thanks to John Bally, Patrice Beaudoin, Jonathan Gagn\'{e}, Maryvonne G\'{e}rin and Sarah Sadavoy for enlightening discussions. We would like to thank the referee for their suggestions, which helped improve the discussion presented in this paper. This study has been achieved thanks to the support of the staff at the Joint Astronomy Centre and the members of the JCMT Gould Belt Survey team. This work would not have been possible without the support of the National Sciences and Engineering Research Council of Canada. This research has made use of the SIMBAD database, operated at CDS, Strasbourg, France.

\appendix
\section{The JCMT Gould Belt Survey Team}

The full members of the JCMT Gould Belt Survey consortium (July 2015) are: 

P. Bastien, D.S. Berry, D. Bresnahan, H. Broekhoven-Fiene, J. Buckle, H. Butner, M. Chen, A. Chrysostomou, S. Coud\'{e}, M.J. Currie, C.J. Davis, J. Di Francesco, E. Drabek-Maunder, A. Duarte-Cabral, M. Fich, J. Fiege, P. Friberg, R. Friesen, G.A. Fuller, S. Graves, J. Greaves, J. Gregson, J. Hatchell, M.R. Hogerheijde, W. Holland, T. Jenness, D. Johnstone, G. Joncas, H. Kirk, J.M. Kirk, L.B.G. Knee, S. Mairs, K. Marsh, B.C. Matthews, G. Moriarty-Schieven, J.C. Mottram, C. Mowat, K. Pattle, J. Rawlings, J. Richer, D. Robertson, E. Rosolowsky, D. Rumble, S. Sadavoy, C. Salji, H. Thomas, N. Tothill, S. Viti, D. Ward-Thompson, G.J. White, J. Wouterloot, J. Yates, and M. Zhu.

\section{Notes on Individual Sources}

Several sources listed in Tables \ref{tab:01} and \ref{tab:02} have also been linked to Spitzer-identified young stellar objects from the \citet{bib:19} catalogue. For completeness, this section lists known objects found in the SIMBAD database within $14.6\arcsec$ of these SCUBA-2 submillimetre sources \citep{bib:46}. These associated objects are listed in Roman numerals under each source presented in the following subsections.

\subsection{Contaminated Sample}

Even though the sources in the contaminated sample (see Table \ref{tab:01}) were identified based on their strong $^{12}$CO J=3-2 line contamination, there are young stars within one 850 $\mu$m effective beam-width ($14.6\arcsec$) of their central position. These are typical variable stars, some of them with significant reddening. There are also two Herbig-Haro emission jets in the surroundings of the contaminated sources. 

\subsubsection{Source C14}
\begin{enumerate}
\item V* V2560 Ori is an Orion-type variable star with a spectral type of M6 located $13.2\arcsec$ mostly south of C14 with $V=19.2$ and $K=11.7$;
\item 2MASS J05353920-0516355 is an irregular-type variable star located $14.08\arcsec$ north-east of C14 with $I=17.4$ and $K=11.4$.
\end{enumerate}

\subsubsection{Source C22}
\begin{enumerate}
\item MGM2012 1539 is an irregular-type variable star with a spectral type of M5.5 located $4.2\arcsec$ north-west of C22 with $V=18.6$ and $K=11.9$;
\item V* V1514 Ori is an Orion-type variable star with a spectral type M4.5e located $8.9\arcsec$ south-west of C22 with $V=17.9$ and $K=10.2$;
\item V* MY Ori (Parenago 1963) is an Orion-type variable star with a spectral type M5e located $9.8 \arcsec$ south-east of C22 with $V=16.2$ and $K=10.2$;
\item HH 561 is an Herbig-Haro object located $10.3\arcsec$ south-east of C22. 
\end{enumerate}

\subsubsection{Source C23}
\begin{enumerate}
\item TKK 544 is an infrared source located $2.7\arcsec$ north-west of C23 with $J=13.3$ and $K=11.1$, and it is somewhat nebulous on the 2MASS image;
\item 2MASS J05351649-0506003 (MGM2012 2363) is a pre-main sequence star also located $2.7\arcsec$ north-west of C23 with $J=16.6$ and $K=13.6$;
\item JBV2003 SK1-OMC3 is a submillimetre source identified with SCUBA by \citet{bib:05} and whose peak is located $5.6\arcsec$ east of C23;
\item HH 357 is an Herbig-Haro object located $11.2\arcsec$ south-west of C23.
\end{enumerate}

\subsubsection{Source C24}
\begin{enumerate}
\item 2MASS J05350553-0551540 is a young stellar object located $4.8\arcsec$ south-east of C24 with $H=16.0$ and $K=13.9$, and it has a nebulous ''tail'' on the 2MASS image.
\end{enumerate}

\subsection{Reference Sample}
The reference sample (see Table \ref{tab:02}) consists of sumillimetre clumps found in OMC-4 and listed in \citet{bib:18}. They were selected as a control group for the contaminated sample because of their low $^{12}$CO J=3-2 line contamination in the 850 $\mu$m band.

\subsubsection{Source 01}
\begin{enumerate}
\item ISOY J053444.05-054125.7 is a young stellar object located $2.2\arcsec$ west of source 01;
\item JCMTSF J053443.8-054126, a submillimetre source identified with SCUBA, is a young stellar object candidate listed in \citet{bib:39} and whose peak is located $6.1\arcsec$ west of source 01. 
\end{enumerate}

\subsubsection{Source 02}
\begin{enumerate}
\item MGM2012 1216 is a pre-main sequence star located $13.0\arcsec$ north-east of source 02;
\item JCMTSE J053449.8-054614 is a submillimetre source identified with SCUBA, listed in \citet{bib:39} and whose peak is located $13.0\arcsec$ mostly west of source 02.
\end{enumerate}

\subsubsection{Source 24}
\begin{enumerate}
\item MGM2012 1400 is a young stellar object located $3.1\arcsec$ of source 24;
\item JCMTSF J053507.9-053556, a submillimetre source identified with SCUBA, is a young stellar object candidate listed in \citet{bib:39} and whose peak is located $6.8\arcsec$ mostly west of source 24;
\item 2MASS J05350800-0535537 is a star located $6.8\arcsec$ north-west of source 24 with $I=21.7$ and $K=15.2$.
\end{enumerate}

\subsubsection{Source 39}
\begin{enumerate}
\item V* V2260 Ori is an Orion-type variable star with a spectral type K4-M0 located $4.8\arcsec$ west of source 39 with $V=19.1$ and $K=9.7$; 
\item NW2007 OrionAN-0535149-53307 is a young stellar object candidate listed in \citet{bib:47} and located $11.3\arcsec$ west of source 39.
\end{enumerate}

\label{lastpage}
\end{document}